\documentclass[11pt]{article}

\usepackage{graphicx}
\usepackage{amssymb}
\usepackage{rotating}
\usepackage[greek,english]{babel}
\usepackage{graphicx}
\usepackage{caption}
\usepackage{lscape}
\usepackage{rotating}
\usepackage{epstopdf}
\usepackage{algorithm}
\usepackage{color}
\usepackage{graphicx,hyperref}
\usepackage{mathtools}
\newcommand{\bfI}{\mbox{\boldmath $I$}}

\newcommand{\bftheta}{\mbox{\boldmath $\theta$}}
\newcommand{\bflambda}{\mbox{\boldmath $\lambda$}}

\newcommand{\bfmu}{\mbox{\boldmath $\mu$}}

\newcommand{\bfbeta}{\mbox{\boldmath $\beta$}}

\newcommand{\bfm}{\mbox{\boldmath $m$}}

\newcommand{\bfy}{\mbox{\boldmath $y$}}

\newcommand{\bfp}{\mbox{\boldmath $p$}}

\newcommand{\bfphi}{\mbox{\boldmath $\phi$}}
\newcommand{\bfgamma}{\mbox{\boldmath $\gamma$}}

\newcommand{\bfw}{\mbox{\boldmath $w$}}
\newcommand{\bfv}{\mbox{\boldmath $v$}}

\newcommand{\bfn}{\mbox{\boldmath $n$}}
\newcommand{\bft}{\mbox{\boldmath $t$}}
\newcommand{\bfzeta}{\mbox{\boldmath $\zeta$}}
\newcommand{\bfT}{\mbox{\boldmath $T$}}

\begin{document}

\parindent0mm \parskip0.6cm

%\begin{center}
{\LARGE {\bf On the correspondence from Bayesian log-linear modelling to logistic regression modelling with $g$-priors} }

\begin{large}

Michail Papathomas

\end{large}

{\it School of Mathematics and Statistics, University of St Andrews, United Kingdom } \\
M.Papathomas@st-andrews.ac.uk

%\end{center}

\vspace{0.5cm}

{\bf Abstract:} Consider a set of categorical variables where at least one of them is binary. The log-linear model that describes the counts in the resulting contingency table implies a specific logistic regression model, with the binary variable as the outcome. Within the Bayesian framework, the $g$-prior and mixtures of $g$-priors are commonly assigned to the parameters of a generalized linear model. We prove that assigning a $g$-prior (or a mixture of $g$-priors) to the parameters of a certain log-linear model designates a $g$-prior (or a mixture of $g$-priors) on the parameters of the corresponding logistic regression. By deriving an asymptotic result, and with numerical illustrations, we demonstrate that when a $g$-prior is adopted, this correspondence extends to the posterior distribution of the model parameters. Thus, it is valid to translate inferences from fitting a log-linear model to inferences within the logistic regression framework, with regard to the presence of main effects and interaction terms.


\section{Introduction}

Consider observations $\bfv=\{v_1,\dots,v_n\}$, parameters $\bftheta=\{\theta_1,\dots,\theta_n\}$, and known quantities or nuisance parameters $\bfphi=\{\phi_1,\dots,\phi_n \}$. Following standard notation, $v_i$, $i=1,\dots,n$, follows a distribution that is a member of the exponential family when its probability function can be written as,
\[
f(v_i | \theta_i, \phi_i) = \mbox{exp} \left\{ \frac{w_i}{\phi_i} \left[v_i \theta_i - b(\theta_i)\right] +c(v_i,\phi_i)
\right\},
\]
where, $\bfw=\{w_1,\dots,w_n \}$ are known weights, and $\phi_i$ is described as the dispersion or scale parameter. With regard to first and second order moments,  
$\mu_i\equiv E(v_i)=b^{'}(\theta_i)$ and $\mbox{Var}(v_i)=\frac{w_i}{\phi_i} b^{''}(\theta_i)$. The variance function is defined as $V(\mu_i) = b^{''}(\theta_i)$. A generalized linear model relates 
$\bfmu=\{\mu_1,\dots,\mu_n \}$ to covariates by setting $\zeta(\bfmu)=X_d \bfgamma$, where $\zeta$ denotes the link function,  $X_d$ the covariate design matrix and $\bfgamma$ a vector of parameters. For a single $\mu_i$, we write $\zeta(\mu_i)=X_{d(i)} \bfgamma$, where $X_{d(i)}$ denotes the $i-th$ row of $X_d$. So, $\zeta$ is defined as a vector function $\zeta \equiv \{\zeta_1,...,\zeta_n \}$ with $n$ elements. 

Denote with $\mathcal{P}$ a finite set of $P$ categorical variables. Observations from $\mathcal{P}$ can be arranged as counts in a $P$-way contingency table. Denote the cell counts as $n_i$,  $i=1,\dots,n_{ll}$. We use the `$ll$' indicator to allude to the log-linear model that will describe these counts. A Poisson distribution is assumed for the counts so that $E(n_i)=\mu_i$. A Poisson log-linear interaction model $\mbox{log}(\bfmu) = X_{ll} \bflambda$ is a generalized linear model that relates the expected counts to $\mathcal{P}$. Assuming that one of the categorical variables, denoted with $Y$, is binary, a logistic regression can also be fitted with $Y$ as the outcome, and all or some of the remaining $P-1$ variables as covariates. We write, $\mbox{logit}(\bfp) = X_{lt} \bfbeta$, $\bfp=(p_1,\dots,p_{n_{lt}})$, using the `$lt$' indicator for the logistic model. Here, $p_i$ denotes the conditional probability that $Y=1$ given covariates $X_{lt(i)}$, and $\bfbeta$ is a vector of parameters. 

Within the Bayesian framework, a prior distribution $f(\bfgamma)$ is assigned to the parameters of the log-linear or logistic regression model. This can be an informative prior that incorporates prior information on the magnitude of the effect of the different covariates or interactions. Eliciting such a prior distribution is not straightforward, especially for the coefficients of interaction terms (Consonni and Veronese 2008). Typically, lack of information for the parameters of a generalized linear model leads to a relatively flat but proper prior distribution, so that model determination based on Bayes factors is  valid (O'Hagan 1995). A very popular choice among Bayesian statisticians is the $g$-prior or a mixture of $g$-priors, described in detail in Section 2. These are flexible priors designed to carry very little information so that inferences are driven by the observed data. See, for example, Wang and George (2007), Saban\`{e}s Bov\`{e} and Held (2011), Overstall and King (2014a;2014b) and Mukhopadhyay and Samantha (2016). This type of prior was first proposed by Zellner (1986) for general linear models. In this context, it is known as Zellner's $g$-prior. Other priors have been proposed, especially for analyses where the focus is on model comparison and variable selection. For example, Jeffreys prior (Liang et al. 2008), the generalized hyper-$g$ prior (Saban\`{e}s Bov\`{e} and Held 2011), and the expected-posterior priors and power-expected-posterior priors (Fouskakis et al. 2015). Our manuscript concerns the $g$-prior and mixture of $g$-priors. After data are collected, the prior $f(\bfgamma)$ is updated to the posterior distribution $f(\bfgamma|\mbox{Data})$ via the conditional probability formula and Bayes Theorem, so that,
\[
f(\bfgamma|\mbox{Data}) = \frac{f(\mbox{Data}|\bfgamma) p(\bfgamma)}{f(\mbox{Data})}. 
\]
For the prior distributions discussed above, closed form expressions for the posterior distribution $f(\bfgamma|\mbox{Data})$ do not exist. The posterior is typically calculated using Markov chain Monte Carlo stochastic simulation, or Normal approximations (O'Hagan and Forster 2004). 

It is known (Agresti 2002) that when $\mathcal{P}$ contains a binary $Y$, a log-linear model $\mbox{log}(\bfmu) = X_{ll} \bflambda$ implies a specific logistic regression model with parameters $\bfbeta$ defined uniquely by $\bflambda$.  The logistic regression model for the conditional odds ratios for $Y$ implies an equivalent log-linear model with arbitrary interaction terms between the covariates in the logistic regression, plus arbitrary main effects for these covariates. 
We provide a simple example to illustrate this result and clarify additional notation. Assume three categorical variables $X,Y$, and $Z$, with $Y$ binary. Let $i,j,k$ be integer indices that describe the level of $X,Y$ and $Z$ respectively. For instance, as $Y$ is binary, $j=0,1$. Consider the log-linear model,
\[
\mbox{log}(\mu_{ijk})=\lambda + \lambda_{i}^{X} + \lambda_{j}^{Y} + \lambda_{k}^{Z} + \lambda_{ij}^{XY} + \lambda_{ik}^{XZ} + \lambda_{jk}^{YZ}, \tag{M1}
\]
where the superscript denotes the main effect or interaction term. The corresponding logistic regression model for the conditional odds ratios for $Y$ is derived as follows, 
\begin{eqnarray}
\mbox{log} \left( \frac{P(Y=1 | X,Z)}{P(Y=0 | X,Z)} \right) &=& 
\mbox{log} \left( \frac{P(Y=1,X,Z)}{P(Y=0,X,Z)} \right) 
%= \mbox{log} \left( \frac{\mu_{i0k}}{\mu_{i1k}} \right) 
\nonumber \\
&=& \mbox{log}(\mu_{i1k}) - \mbox{log}(\mu_{i0k}) \nonumber \\ 
&=& \lambda_{1}^{Y} - \lambda_{0}^{Y} + \lambda_{i1}^{XY} - \lambda_{i0}^{XY} 
+ \lambda_{1k}^{YZ} - \lambda_{0k}^{YZ}. \nonumber 
\end{eqnarray}
This is a logistic regression with parameters,  $\bfbeta=(\beta,\beta_{i}^{X},\beta_{k}^{Z})$, so that, $\beta=\lambda_{1}^{Y} - \lambda_{0}^{Y}$, $\beta_{i}^{X}=\lambda_{i1}^{XY} - \lambda_{i0}^{XY}$, and $\beta_{k}^{Z}=\lambda_{1k}^{YZ} - \lambda_{0k}^{YZ}$. Considering identifiability corner point constraints, all elements in $\bflambda$ with a zero subscript are set to zero. Then, $\beta=\lambda_{1}^{Y}$, $\beta_{i}^{X}=\lambda_{i1}^{XY} $ and $\beta_{k}^{Z}=\lambda_{1k}^{YZ}$. This scales in a straightforward manner to larger log-linear models. For instance, if (M1) contained the three-way interaction $XYZ$, then the corresponding logistic regression model would contain the $XZ$ interaction, so that, $\beta_{ik}^{XZ}=\lambda_{i1k}^{XYZ} - \lambda_{i0k}^{XYZ}$, and under corner point constraints, $\beta_{ik}^{XZ}= \lambda_{i1k}^{XYZ}$. If a factor does not interact with $Y$ in the log-linear model, then this factor disappears from the corresponding logistic regression model. 
To demonstrate that the correspondence between log-linear and logistic models is not bijective, it is straightforward to show that, for example, the log-linear model, $\mbox{log}(\mu_{ijk})=\lambda + \lambda_{i}^{X} + \lambda_{j}^{Y} + \lambda_{k}^{Z} + \lambda_{ij}^{XY} + \lambda_{jk}^{YZ}$, implies the same logistic regression as (M1). 
More generally, the relation between $\bfbeta$ and $\bflambda$ can be described as $\bfbeta=\bfT \bflambda$, where $\bfT$ is an incidence matrix (Bapat 2011). In the context of this manuscript, matrix $\bfT$ has one row for each element of $\bfbeta$, and one column for each element of $\bflambda$. The elements of $\bfT$ are zero, except in the case where the element of $\bfbeta$ is defined by the corresponding element of $\bflambda$. The number of rows of $\bfT$ cannot be greater than the number of columns. To simplify the analysis and notation, for the remainder of this manuscript we consider models specified under corner point constraints. Then, every logistic regression model parameter is defined uniquely by the corresponding log-linear model parameter, and the correspondence from a log-linear to a logistic regression model is direct. 

The contribution of our manuscript is two-fold. First, Theorem 1 states that assigning to $\bflambda$ the $g$-prior that is specific to log-linear modelling, implies the $g$-prior specific to logistic modelling on the parameters $\bfbeta$ of the corresponding logistic regression. The log-linear model has to be the largest model that corresponds to the logistic regression, i.e. the model that contains all possible interaction terms between the categorical factors in $\mathcal{P} \setminus \{ Y \}$. Second, under the reasonable assumption that an investigator who chooses a $g$-prior for $\bflambda$ would also choose a $g$-prior for $\bfbeta$ if they were to fit a logistic regression directly, inferences on the parameters of a log-linear model translate to inferences on the parameters of the corresponding logistic regression. Closed form expressions for the posterior distributions do not exist. Wang and George (2007) utilize the Laplace approximation for generalized linear models, focusing on the approximation of the marginal likelihood for the purpose of variable selection. Theorem 2 shows that, asymptotically, the matching between the prior distributions of the corresponding parameters extends to the posterior distributions. It is then demonstrated by numerical illustrations that the presence or absence of interaction terms in the log-linear model can inform on the relation between the binary $Y$ and the other variables as described by logistic regression. For example, assume that after fitting a specific log-linear model, the credible interval for an element of $\bflambda$  contains zero. When fitting the corresponding logistic regression model, the investigator will anticipate that the credible interval for the corresponding element of $\bfbeta$ will also contain zero. {\it Importantly}, for this translation to hold, it is essential that the prior distribution for $\bfbeta$ implied by the prior on $\bflambda$ is the same to the distribution the investigator would assign to $\bfbeta$ if they were to fit the logistic model directly. If the implied prior on $\bfbeta$ is not the same as a directly assigned prior then, with regard to $\bfbeta$, the correspondence from the Bayesian log-linear analysis to the logistic one becomes dubious. In both illustrations in Section 4, we observe that the credible intervals of the corresponding $\bflambda$ and $\bfbeta$ parameters are virtually identical considering simulation error. 

In Section 2, we provide the definition of the $g$-prior and mixtures of $g$-priors, and describe how the $g$-prior is derived for log-linear and logistic regression models. Section 3, contains the main contributions in this manuscript. In Section 4, the correspondence from a log-linear to a  logistic regression model is illustrated using simulated and real data. We conclude with a discussion.   

\section{The $g$-prior and mixtures of $g$-priors}

A $g$-prior for the parameters $\bfgamma$ of a generalized linear model is a multivariate Normal distribution $N(\bfm_{\gamma},g \Sigma_{\gamma})$, constructed so that the prior variance is a multiple of the inverse Fisher information matrix by a scalar $g$. See Liang et al. (2008) for a discussion on the choice of $g$. In accordance with Ntzoufras et al. (2003) and Ntzoufras (2009), the $g$-prior for the parameters of log-linear and logistic regression models is specified so that, 
$\bfm_{\gamma}=(m_{\gamma_1},0,\dots,0)^{\top}$, 
where $m_{\gamma_{1}}$ corresponds to the intercept and can be non-zero, and,
\[
\Sigma_{\gamma} = V(m^{*}) \zeta^{'}(m^{*})^{2} [(X_{d}^{\top} \mbox{diag}(\frac{1}{\phi_i}) X_{d}]^{-1},
\]
where $\mbox{diag}(1/\phi_i)$ denotes a diagonal $n\times n$ matrix with non-zero elements $1/\phi_i$, and $m^{*}=\zeta^{-1}(m_{\gamma_1})$.

The unit information prior is a special case of the $g$-prior, obtained by setting $g=N$, where $N$ denotes the total number of observations. It is constructed so that the information contained in the prior is equal to the amount of information in a single observation (Kass and Wasserman 1995). 
Assuming that $g$ is a random variable, with prior $f(g)$, leads to a mixture of $g$-priors, so that, 
\[
\bfgamma | g \sim N(\bfm_{\gamma},g \Sigma_{\gamma}), \mbox{ } g \sim f(g).
\]
Mixtures of $g$-priors are also called hyper-$g$ priors (Saban\`{e}s Bov\`{e} and Held 2011). 

\underline{Log-linear regression:} Consider counts $n_i$  $i=1,\dots,n_{ll}$. Now, $N=\sum_{i=1}^{n_{ll}} n_i$, and,
\[
f(n_i | \mu_i) = \frac{e^{-\mu_i} \mu_{i}^{n_i}}{n_i !} ,
\] 
with $\theta_i = \mbox{log}(\mu_i)$,
$b(\theta_i)=e^{\theta_i}$ and $c(n_i,\phi_i)=\mbox{-log} (n_i !)$. 
Also, $w_i \phi_{i}^{-1}=1$, so that $w_i=1$ implies $\phi_i=1$. Note that, 
\[
\mu_i=b^{'}(\theta_i)=e^{\theta_i}, \mbox{ }
\mbox{Var}(n_i)=\phi_i w_{i}^{-1} b^{''}(\theta)=e^{\theta_i}, \mbox{ } \mbox{ and } V(\mu_i)=\mu_i. 
\]
For the log-linear model, $\mbox{log}(\bfmu) = X_{ll} \bflambda$, and 
$\zeta(\mu_i) = \mbox{log}(\mu_i)$ so that $\zeta^{'}(\mu_i)=\mu_{i}^{-1}$. 
The $g$-prior is constructed as $N(\bfm_{\lambda},g \Sigma_{\lambda})$, where,
$\bfm_{\lambda}=(\mbox{log}(\bar{n}),0,\dots,0)$. Here,  $\bar{n}$ denotes the average cell count. The prior mean for the log-linear model intercept is also often set to zero (Dellaportas et al. 2012). (Note that altering the prior mean for the log-linear model intercept does not affect the validity of the theoretical results in Section 3. This is straightforward to deduce from the proof of Theorem 1 given in the Appendix, as the prior mean for the log-linear intercept does not affect the implied distribution of the logistic regression parameters.) In addition, 
\[
\Sigma_{\lambda} = \bar{n} \frac{1}{(\bar{n})^2} 
(X_{ll}^{\top} X_{ll})^{-1} 
= \frac{1}{\bar{n}} (X_{ll}^{\top} X_{ll})^{-1} 
= \frac{n_{ll}}{N} (X_{ll}^{\top} X_{ll})^{-1}.
\]

\underline{Logistic regression:} Assume that $y_i$, $i=1,\dots,n_{lt}$, is the proportion of successes out of $t_i$ trials. Now, $N=\sum_{i=1}^{n_{lt}} t_i$, and,
\[
f(t_i y_i | p_i) = {t_i \choose t_i y_i} p_i^{t_i y_i} (1-p_i)^{t_i - t_i y_i},
\] 
where $\theta_i=\mbox{logit}(p_i)$,  $b(\theta_i)=\mbox{log}(1+e^{\theta_i})$, and  $c(y_i,\phi_i)=\mbox{log} {t_i \choose t_i y_i}$. Also, $w_i \phi_{i}^{-1}=t_i$, so that $w_i=1$ implies  $\phi_i=t_{i}^{-1}$. Note that, 
\[
E(y_i)=b^{'}(\theta_i)=\frac{e^{\theta_i}}{1+e^{\theta_i}}=p_i, \mbox{ } 
\mbox{Var}(y_i)=\frac{\phi_i}{w_i} b^{''}(\theta_i)=\frac{1}{t_i} \frac{e^{\theta_i}}{(1+e^{\theta_i})^2}=\frac{p_i(1-p_i)}{t_i}, 
\]
and,
\[
V(p_i)=p_i(1-p_i). 
\]
The logistic regression model is defined as $\mbox{logit}(\bfp) = X_{lt} \bfbeta$, so that $X_{lt}$ is a $n_{lt} \times n_{\beta}$ design matrix, and  
$\zeta(p_i) = \mbox{logit}(p_i)$ so that $\zeta^{'}(p_i)=[p_i(1-p_i)]^{-1}$.
The $g$-prior is $N(\bfm_{\beta},g \Sigma_{\beta})$, where, $\bfm_{\beta}=(0,0,\dots,0)$, and, 
\begin{eqnarray}
\Sigma_{\beta} =  p^{*} (1-p^{*}) \frac{1}{[p^{*} (1-p^{*})]^2} 
[X_{lt}^{\top} \mbox{diag}(t_i) X_{lt}]^{-1} = \frac{1}{0.25} [X_{lt}^{\top} \mbox{diag}(t_i) X_{lt}]^{-1}. \nonumber
\end{eqnarray}
Here, $p^{*}$ corresponds to $m^{*}$ in the general definition of the $g$-prior at the start of this Section, so that $p^{*}=\zeta^{-1}(m_{\gamma_{1}})$, where $m_{\gamma_{1}}$ is the first element of $\bfm_{\beta}$ which is zero. Thus, we obtain that $p^{*}=e^{0}/(e^{0}+1)=0.5$. By approximating each $t_i$ with the average number of trials $\bar{t}$, as suggested by Ntzoufras et al. (2003), 
\[
\Sigma_{\beta} \simeq 4 \frac{1}{\bar{t}} (X_{lt}^{\top} X_{lt})^{-1} = 
4 \frac{n_{lt}}{\sum_{i=1}^{n_{lt}} t_i} (X_{lt}^{\top} X_{lt})^{-1} = 
4 \frac{n_{lt}}{N} (X_{lt}^{\top} X_{lt})^{-1}.
\]

\section{Correspondence from log-linear to logistic regression models}

Consider a set of categorical variables $\mathcal{P}$ that includes a binary variable $Y$. Assume a log-linear model that, in addition to the terms that involve $Y$, contains all possible interaction terms between the categorical factors in $\mathcal{P} \setminus \{ Y \}$. We show that, given that a $g$-prior is assigned to the log-linear model parameters $\bflambda$, the implied prior for $\bfbeta$ is a $g$-prior for logistic regression models, i.e. the one that would be assigned if the investigator considered the logistic regression model directly. 

{\bf Theorem 1:} Assume a $g$-prior $\bflambda \sim N(\bfm_{\lambda}, g \Sigma_{\lambda})$ on the parameters of a log-linear model $\mbox{log}(\bfmu) = X_{ll} \bflambda$, that contains all possible interaction terms between the categorical factors in $\mathcal{P} \setminus \{ Y \}$. This prior implies a $g$-prior $N(\bfm_{\beta}, g \Sigma_{\beta})$ for the parameters $\bfbeta$ of the corresponding logistic regression $\mbox{logit}(\bfp) = X_{lt} \bfbeta$. 

{\it Proof:} The proof is based on rearranging the rows and columns of $X_{ll}$, and partitioning so that one part of $X_{ll}$ consists of the logistic design matrix $X_{lt}$, or replications of $X_{lt}$. We then show that the prior mean and variance of the elements of $\bflambda$ that correspond to $\bfbeta$ is the prior that would be assigned to $\bfbeta$ if the logistic regression was fitted directly.  The complete proof is given in the Appendix.  

{\bf Corollary 1:} A unit information prior $\bflambda \sim N(\bfm_{\lambda}, N \Sigma_{\lambda})$ implies a unit information prior $N(\bfm_{\beta}, N \Sigma_{\beta})$ for the parameters $\bfbeta$ of the corresponding logistic regression. 

Corollary 1 follows directly from Theorem 1 by setting $g=N$. The following Corollary concerns mixtures of $g$-priors. It is implicitly assumed that the investigator would adopt the same prior density $f(g)$ for both modelling approaches.  

{\bf Corollary 2:} A mixture of $g$-priors so that $\bflambda | g \sim N(\bfm_{\lambda}, g \Sigma_{\lambda})$, $g\sim f(g)$, implies a mixture of $g$-priors for the parameters $\bfbeta$ of the corresponding logistic regression, so that $\bfbeta | g \sim N(\bfm_{\beta}, g \Sigma_{\beta})$, $g\sim f(g)$. 

This also follows from Theorem 1, which states that when $\bflambda | g \sim N(\bfm_{\lambda}, g \Sigma_{\lambda})$, the conditional prior for $\bfbeta$ is $\bfbeta | g \sim N(\bfm_{\beta}, g \Sigma_{\beta})$. 

%\underline{Considering a flat prior on the intercept:}
When the $g$-prior is utilized, it is common to assign a locally uniform Jeffreys prior ($\propto 1$) on the intercept, after the covariate columns of the design matrix have been centered to ensure orthogonality with the intercept (Liang et al., 2008). If one decides to adopt the approach where a flat prior is assigned to the intercept in both log-linear and logistic formulations, the correspondence between log-linear and logistic regression breaks, but only with regard to the intercept of the logistic regression. The prior on the log-linear intercept does not have a bearing on the implied prior for the logistic regression parameters, because the log-linear intercept does not contribute to the formation of the logistic regression parameters, as described in Section 1. After assigning a flat prior on the intercept of the log-linear model, all $\bfbeta$ parameters (including the intercept) are still Normal as linear combinations of Normal random variables, and the distribution of $\bfbeta$ is the one given by Theorem 1. For details see the additional material in the proof of Theorem 1 in the Appendix. For an illustration, see Table 3 in Section 4.2. 

Closed form expressions for the posterior distribution of the parameters of a generalized linear model do not exist. However, it is known (O'Hagan and Forster 2004) that a Normal approximation applies. Consider a $g$-prior for the parameters $\bfgamma$ of the generalized linear model, $\zeta(\bfmu)=X_d \bfgamma$, so that, for fixed $g$, 
\[
\bfgamma \sim N(\bfm_{\gamma},g \Sigma_{\gamma}).
\]

Given observations $\bfv =\{ v_1,\hdots,v_n \}$, the posterior distribution of 
$\gamma$ is approximated by a Normal density, so that, 
\begin{eqnarray}
\bfgamma | \bfv 
\sim N([g^{-1} \Sigma_{\gamma}^{-1} + {\cal I}(\hat{\bfgamma})]^{-1} \times 
[g^{-1} \Sigma_{\gamma}^{-1} \bfm_{\gamma} + {\cal I}(\hat{\bfgamma}) \hat{\bfgamma}] ,
  [g^{-1} \Sigma_{\gamma}^{-1} + {\cal I}(\hat{\bfgamma})]^{-1}  ).
\end{eqnarray}
Here, $\hat{\bfgamma}$ is the maximum likelihood estimate of ${\bfgamma}$, and ${\cal I}(\hat{\bfgamma})$ is the information matrix $X_d^{\top} {\cal V} X_d$. For the log-linear model, the diagonal matrix ${\cal V}$ (denoted by ${\cal V}_{log-linear}$), has diagonal elements $\mbox{exp}\{ X_{ll(i)} \hat{\bflambda} \}$, $i=1,\dots,n_{ll}$. 
When the logistic regression is fitted, ${\cal V}_{logistic}$ has diagonal elements $t_i \mbox{exp}\{ X_{lt(i)} \hat{\bfbeta} \} \mbox{exp}\{1 + X_{lt(i)} \hat{\bfbeta} \}^{-2}$, $i=1,\dots,n_{lt}$. Within the Bayesian framework, when fitting a generalized linear model, a large sample $(n \rightarrow \infty)$ will swamp the prior distribution, rendering it irrelevant for deriving posterior inferences (O'Hagan and Forster 2004). In practice, this can be true even for moderate sample sizes (say, of order $10^2$ or larger), especially when the prior is not informative, which is typically the case with $g$-priors. 

{\bf Theorem 2:} Consider a $g$-prior $\bflambda \sim N(\bfm_{\lambda}, g \Sigma_{\lambda})$ on the parameters of a log-linear model $\mbox{log}(\bfmu) = X_{ll} \bflambda$, that contains all possible interaction terms between the categorical factors in $\mathcal{P} \setminus \{ Y \}$. Consider also the analogous $g$-prior $N(\bfm_{\beta}, g \Sigma_{\beta})$ for the parameters $\bfbeta$ of the corresponding logistic regression $\mbox{logit}(\bfp) = X_{lt} \bfbeta$. For fixed $g$, and for a large sample, the posterior distribution of $\bfbeta$, as given in (1), is approximately equal to the posterior distribution of the elements of $\bflambda$ that correspond to $\bfbeta$.

{\it Proof:} A partitioning similar to the one adopted for the proof of Theorem 1 is utilized. First, we show that, asymptotically, the posterior variance of $\bfbeta$ is identical to the posterior variance of the elements of $\bflambda$ that correspond to $\bfbeta$. Then, we do the same for the posterior means. The proof is based on the assumption that for a large sample the contribution of the prior in deriving the posterior moments can be ignored. A standard result utilized in the proof is that, asymptotically, the Binomial distribution for a data point can be approximated by a Poisson distribution. The complete proof is given in the Appendix. 

In the next Section, we demonstrate with numerical illustrations that, for fixed $g$, the correspondence between the priors extends to posterior distributions, so that the posterior distribution of the logistic regression parameters matches the one of the corresponding log-linear model parameters. This is true even for relatively moderate sample sizes $N$, say a few hundred, and for standard choices of $g$ such as $g=N$. 

\section{Illustrations}

Unit information priors were adopted for the model parameters ($g=N$). The size of the burn-in sample was $10^4$, followed by $5\times 10^5$ iterations. 

\subsection{A simulation study}

We simulate data from 1000 subjects, on six binary variables $\{Y,A,B,C,D,E\}$. Probabilities that correspond to the cells of the $2^6$ contingency table are generated in accordance with the log-linear model, $\mbox{log}(\bfmu)=YAB+YCD+YE$. Adopting the notation in Agresti (2002), a single letter denotes the presence of a main effect, two letter terms denote the presence of the implied first-order interaction and so on and so forth. The presence of an interaction between a set of variables implies the presence of all lower order interactions plus main effects for that set. Cell counts are simulated according to the generated cell probabilities. Parameter values and the design matrix of the log-linear model used to generate the cell probabilities are given in the Supplemental material, Section S2. 

We fit to the simulated data the log-linear model,
\[
\mbox{log}(\bfmu)=YAB+YCD+YE+ABCDE. \hspace{2cm} \tag{M2}
\]
According to the discussion and results in Sections 1 and 3, the logistic regression where $Y$ is treated as the outcome should only contain the first-order interactions $AB$ and $CD$ plus the main effect for $E$,
\[
\mbox{logit}(\bfp) = AB+CD+E. \tag{M3}
\] 
In Table 1, we present credible intervals (CI) for the parameters of (M3) and the relevant parameters of (M2). The CIs for the corresponding $\bflambda$ and $\bfbeta$ parameters are almost identical, considering simulation error. For example, the CI for $\lambda_{1,1,1}^{YCD}$ is $(-2.01,-0.85)$, whilst the CI for $\beta_{1,1}^{CD}$ is $(-2.00,-0.84)$. 

In Table 2, we present minimum,  maximum and quantile values for the $t_i$ observations, for each logistic regression shown in Table 1. It is clearly demonstrated that the simulated data do not represent balanced Binomial experiments where $t_i=\bar{t}$. The credible intervals shown in Table 1 demonstrate that the correspondence studied in this manuscript is very robust to departures from $t_i = \bar{t}$. This is also demonstrated in the real data analysis presented in the next subsection, where the collected data do not represent balanced Binomial experiments when one of the factors is treated as the outcome. In the Supplemental material we present additional analyses on simulated data sets, including results on smaller samples, roughly one quarter the size of the data set analysed in this Section. Inferences on the correspondence between the posterior distributions remain unchanged. 

%In terms of deviance, a logistic regression model is equivalent to the largest possible log-linear model that corresponds to it. In our example, (M3) attains a deviance of 20.178, whilst (M2) produces 33.707. The log-linear model that is equivalent in terms of deviance to (M3) is $\mbox{log}(\bfmu)=ABCDE+YAB+YCD+YE$. 

\subsection{A real data illustration}

Edwards and Havr\'anek (1985) presented a $2^{6}$ contingency table
in which $1841$ men were cross-classified by six binary risk factors 
$\{A,B,C,D,E,F\}$ for
coronary heart disease. the data were also analyzed in Dellaportas and Forster (1999), where the top Hierarchical model was,  
$\mbox{log}(\bfmu)=AC+AD+AE+BC+CE+DE+F$, 
with posterior model probability 0.28. In Table 3, we present CIs for parameters of the log-linear model 
\[AC+AD+AE+BCDEF. \tag{M4}
\]
We also present CIs for the parameters of the corresponding logistic regression model when $A$ is treated as the outcome, 
\[
\mbox{logit}(\bfp) = C+D+E. \tag{M5}
\]

We performed this analysis twice. Once after considering the $g$-priors described in Section 2 ($g=N$), as in the previous illustration, and after adopting a $g$-prior with a locally flat prior for the intercept. Under the $g$-prior described in Section 2, the CIs for the corresponding $\bflambda$ and $\bfbeta$ parameters (including the intercept) are almost identical, considering simulation error. For instance, the CI for both the coefficient of $A$ in the log-linear model and the intercept in the logistic regression is $(-0.59,-0.24)$. 
Under the flat prior for the intercepts, the correspondence breaks down with regard to the intercept in the logistic regression model. The CI for the coefficient of $A$ in the log-linear model is $(-0.59,-0.24)$ whilst the CI for the intercept of the corresponding logistic regression model is $(-0.17,0.02)$. Concurrently, the credible intervals for the coefficients of $C$, $D$ and $E$ in the logistic regression model are almost identical to the corresponding CIs for $AC$, $AD$ and $AE$ in the log-linear model, with differences due to simulation error. 

%Model, $\mbox{logit}(\bfp) = C+D+E$, attains a deviance of 33.513, whilst $(M4)$ produces 64.922, and $(M5)$  65.847. The log-linear model that is equivalent to $\mbox{logit}(\bfp) = C+D+E$ in terms of deviance is $\mbox{log}(\bfmu)=BCDEF+AC+AD+AE$.

\section{Discussion}

The correspondence we investigated is not unexpected, given the results in Agresti (2002) discussed in the Introduction, and also the link between the $g$-prior and Fisher's information matrix (Held et al. 2015), although this link is stronger for general linear models. Our investigation is also related to Consonni and Veronese (2008), where specifying a prior for the parameters of one model, and then transferring this specification to the parameters of another is discussed. Of the four strategies considered in Consonni and Veronese (2008), the one directly linked to our manuscript is `Marginalization', as the derived prior for the parameters of the logistic regression is the one that is the marginal prior of the relevant parameters of the log-linear model. Results on the relation between different statistical models are of interest, as they improve understanding and enhance the models' utility. Often, developments for one modelling framework are not readily available for the other. For example, Papathomas and Richardson (2016) comment on the relation between log-linear modelling and variable selection within clustering, in particular with regard to marginal independence, without examining logistic regression models. 

Our numerical illustrations concern the $g$-prior, where the parameter $g$ is fixed. To further explore the correspondence between the two modelling frameworks, we also considered the two hyper priors that are prominent in Liang et al. (2008). This is the Zellner-Siow prior [IG(0.5,N/2)], and the prior introduced in the aforementioned manuscript in Section 3.2, with the suggested specification $\alpha=3$. Furthermore, the two data sets were analysed after adopting a mixture of $g$-priors such that, $g\sim \mbox{IG}(a_{g},b_{g})$. We considered $a_g = 2+\mbox{mean}(g)^2/\mbox{var}(g)$ and $b_g = \mbox{mean}(g) + \mbox{mean}(g)^3/\mbox{var}(g)$, in accordance with the specified prior moments $\mbox{mean}(g)$ and $\mbox{var}(g)$. We considered distinct Inverse Gamma densities with markedly different expectations and variances, as well as the vague prior $IG(0.1,0.1)$. We observed that the correspondence does not hold exactly when a mixture of $g$-priors is adopted. This seems to be because the posterior distribution for $g$ changes under the two modelling frameworks, something that affects to a small, but noticeable degree, the posterior credible intervals for the model parameters. For more details see the analyses presented in the Supplemental material. 

Theoretical results in this manuscript refer to a specific log-linear model and the corresponding logistic regression model, for a given set of covariates. Therefore, our results should not be misinterpreted as license to readily translate log-linear model selection inferences to inferences concerning logistic regression models.
When performing model selection in a space of log-linear models, the prominent log-linear model describes a certain dependence structure between the categorical factors, including the relation of the binary Y with all other factors. 
The logistic regression that corresponds to the prominent log-linear model describes the dependence structure between Y and the other factors that is supported by the data in accordance with the log-linear analysis. 
Therefore, under reasonable expectation, results from a single log-linear model determination analysis may translate, at the very least, to interesting logistic regressions for any of the binary factors that formed the contingency table. However, the mapping between log-linear and logistic regression model spaces is not bijective. Furthermore, posterior model probabilities depend on the prior on the model space, with various different approaches for defining such a prior discussed in Dellaportas et al. (2012). 
For the simulated data analysed in Section 4.1, log-linear model $YAB+YCD+YE$ has posterior probability 0.98, whilst the posterior probability of the corresponding logistic regression model (M3) is 0.59. Similar results from analysing the real data in Section 4.2, not presented here, also support this note of caution. In all model determination analyses, the Reversible Jump MCMC algorithm proposed in Papathomas et al. (2011) was employed. All possible graphical log-linear models were assumed equally likely a priori, as were all possible logistic graphical models for some given outcome.

\section{Acknowledgements}
The author wishes to thank Professor Petros Dellaportas and Dr Antony Overstall for useful discussions during the preparation of this manuscript. We would also like to thank two Reviewers and the Editors for comments that helped to improve the manuscript. 
%If you'd like to thank anyone, place your comments here
%and remove the percent signs.

\vspace{0.5cm}

\noindent {\bf Appendix}

\vspace{0.1cm}
{\it {\bf Proof of Theorem 1:}} To facilitate the proof, the following notation is introduced. Using the incidence matrix $\bfT$ discussed in Section 1, write the mapping between $\bfbeta$ and $\bflambda$ as $\bfbeta=\bfT \bflambda$, where, 
\[
  \bfT=\left( \begin{array}{l}
    \bflambda_{(1)} \\
    \vdots \\
    \bflambda_{(n_{\lambda_{Y}})} \\
  \end{array}
  \right),
\]
and $\bflambda_{(k)}$, $k=1,\dots,n_{\lambda_{Y}}$, is a vector of zeros with the exception of one element that is equal to one. This element is in the position of the $k$-th $\bflambda$ parameter with a $Y$ in its superscript. With $n_{\lambda_{Y}}$ we denote the number of parameters in $\bflambda$ with a $Y$ in their superscript. This is a more rigorous definition of $\bfT$ compared to the more descriptive definition in Section 1.  
To ease algebraic calculations, and without any loss of generality, rearrange the columns of $\bflambda$, creating a new vector $\bflambda_{r}$, so that $\bfT$ changes accordingly to,
$
  \bfT_{r}=\left( \begin{array}{ll}
    \bfI & \bf0  
  \end{array}
  \right),
$ 
where $\bfI$ is an $n_{\beta}\times n_{\beta}$ identity matrix and $n_{\beta}$ is the number of elements in $\bfbeta$. The rows and columns of $X_{ll}$ are also rearranged accordingly to create $X_{rll}$, so that,
\begin{eqnarray}
  X_{rll}=\left( \begin{array}{ll}
    X_{lt}^{*} & X_{ll-lt} \\
     \bf0 & X_{ll-lt}\\
  \end{array}
  \right) \hspace{5cm} \mbox{(A.1)} \nonumber
\end{eqnarray}

$X_{ll-lt}$ is a square $(n_{ll}/2 \times n_{ll}/2)$ matrix. This is  because we consider the log-linear model that, in addition to the terms that involve $Y$, contains all possible interaction terms between the categorical factors in 
$\mathcal{P} \setminus \{ Y \}$. The number of parameters that correspond to the intercept, main effects and interactions for $\mathcal{P} \setminus \{ Y \}$ is $n_{ll}/2$. 

Denote with $j_1=2$ the number of levels of the binary factor $Y$ that becomes the outcome in the logistic regression model. With $j_2$ to $j_q$, $1\leq q \leq P-1$ denote the number of levels of the $q-1$ factors that are present in the log-linear model but disappear from the logistic regression model as they do not interact with $Y$. Then, $n_{ll}=2\times j_2 \times \hdots \times j_q \times n_{lt}$. When $q=1$, all factors other than $Y$ remain in the logistic regression model as covariates. When $q=P-1$, the corresponding logistic regression model only contains the intercept. For instance, for a $2^P$ contingency table, $n_{ll}=2^q \times n_{lt}$, and for $q=1$, $n_{ll}=2\times n_{lt}$. 
Furthermore, $X_{lt}^{*}$ is a $n_{ll}/2 \times n_{\beta}$ matrix. By rearranging the rows of $X_{rll}$ when necessary, we can write $X_{lt}^{*}$ as, 
$X_{lt}^{*}=(X_{lt}^{\top} X_{lt}^{\top} \hdots X_{lt}^{\top})^{\top}$, where $X_{lt}^{\top}$ is repeated
$ (j_1-1) \times j_2 \times \hdots \times j_q$ times.  For example, for $q=1$, $X_{lt}^{*}=X_{lt}$. For $q=2$, $X_{lt}$ repeats $j_2$ times within $X_{lt}^{*}$. 

We can now write $\bfbeta=\bfT_{r} \bflambda_{r}$. For example, assume the log-linear model (M1) describes a $3\times 2\times 2$ contingency table. Then, $q=1$, and the standard arrangement of the elements of $\bflambda$ would be such that, 
\begin{tiny}
 \[
  X_{ll} = \left( \begin{array}{llllllllll}
  1 & 0 & 0 & 0 & 0 & 0 & 0 & 0 & 0 & 0 \\
  1 & 1 & 0 & 0 & 0 & 0 & 0 & 0 & 0 & 0 \\
  1 & 0 & 1 & 0 & 0 & 0 & 0 & 0 & 0 & 0 \\
  1 & 0 & 0 & 1 & 0 & 0 & 0 & 0 & 0 & 0 \\
  1 & 1 & 0 & 1 & 0 & 1 & 0 & 0 & 0 & 0 \\
  1 & 0 & 1 & 1 & 0 & 0 & 1 & 0 & 0 & 0 \\
  1 & 0 & 0 & 0 & 1 & 0 & 0 & 0 & 0 & 0 \\
  1 & 1 & 0 & 0 & 1 & 0 & 0 & 1 & 0 & 0 \\
  1 & 0 & 1 & 0 & 1 & 0 & 0 & 0 & 1 & 0 \\
  1 & 0 & 0 & 1 & 1 & 0 & 0 & 0 & 0 & 1 \\
  1 & 1 & 0 & 1 & 1 & 1 & 0 & 1 & 0 & 1 \\
  1 & 0 & 1 & 1 & 1 & 0 & 1 & 0 & 1 & 1 
  \end{array}
  \right), \hspace{0.2cm}
  \bflambda   = \left( \begin{array}{l}
  \lambda \\ \lambda_{1}^{X} \\ \lambda_{2}^{X} \\  \lambda_{1}^{Y} \\ \lambda_{1}^{Z} \\ \lambda_{11}^{XY} \\  \lambda_{21}^{XY} \\ \lambda_{11}^{XZ} \\ \lambda_{21}^{XZ} \\     \lambda_{11}^{YZ} 
  \end{array}
  \right), \hspace{0.2cm}
  \bfT = \left( \begin{array}{llllllllll}
  0 & 0 & 0 & 1 & 0 & 0 & 0 & 0 & 0 & 0 \\
  0 & 0 & 0 & 0 & 0 & 1 & 0 & 0 & 0 & 0 \\
  0 & 0 & 0 & 0 & 0 & 0 & 1 & 0 & 0 & 0 \\
  0 & 0 & 0 & 0 & 0 & 0 & 0 & 0 & 0 & 1 
  \end{array}
  \right)
  \]
 \end{tiny}

After rearranging, 
\begin{tiny}
 \[
  X_{rll} = \left( \begin{array}{llllllllll}
  1 & 0 & 0 & 0 & 1 & 0 & 0 & 0 & 0 & 0 \\
  1 & 1 & 0 & 0 & 1 & 1 & 0 & 0 & 0 & 0 \\
  1 & 0 & 1 & 0 & 1 & 0 & 1 & 0 & 0 & 0 \\
  1 & 0 & 0 & 1 & 1 & 0 & 0 & 1 & 0 & 0 \\
  1 & 1 & 0 & 1 & 1 & 1 & 0 & 1 & 1 & 0 \\
  1 & 0 & 1 & 1 & 1 & 0 & 1 & 1 & 0 & 1 \\
  0 & 0 & 0 & 0 & 1 & 0 & 0 & 0 & 0 & 0 \\
  0 & 0 & 0 & 0 & 1 & 1 & 0 & 0 & 0 & 0 \\
  0 & 0 & 0 & 0 & 1 & 0 & 1 & 0 & 0 & 0 \\
  0 & 0 & 0 & 0 & 1 & 0 & 0 & 1 & 0 & 0 \\
  0 & 0 & 0 & 0 & 1 & 1 & 0 & 1 & 1 & 0 \\
  0 & 0 & 0 & 0 & 1 & 0 & 1 & 1 & 0 & 1 
  \end{array}
  \right), \hspace{0.2cm}
  \bflambda_{r}   = \left( \begin{array}{l}
  \lambda_{1}^{Y} \\ \lambda_{11}^{XY} \\ \lambda_{21}^{XY} \\  \lambda_{11}^{YZ} \\ \lambda \\ \lambda_{1}^{X} \\  \lambda_{2}^{X} \\ \lambda_{1}^{Z} \\ \lambda_{11}^{XZ} \\ \lambda_{21}^{XZ} 
  \end{array}
  \right), \hspace{0.2cm}
  \bfT_{r} = \left( \begin{array}{llllllllll}
  1 & 0 & 0 & 0 & 0 & 0 & 0 & 0 & 0 & 0 \\
  0 & 1 & 0 & 0 & 0 & 0 & 0 & 0 & 0 & 0 \\
  0 & 0 & 1 & 0 & 0 & 0 & 0 & 0 & 0 & 0 \\
  0 & 0 & 0 & 1 & 0 & 0 & 0 & 0 & 0 & 0 
  \end{array}
  \right)
  \]
 \end{tiny}

For another example, where $q=2$, consider again model (M1) but now assume that the interaction $YZ$ is not present in the log-linear model. Then, the $Z$ factor will disappear from the corresponding logistic regression model, and after rearranging, 
\begin{tiny}
 \[
  X_{rll} = \left( \begin{array}{lllllllll}
  1 & 0 & 0 & 1 & 0 & 0 & 0 & 0 & 0 \\
  1 & 1 & 0 & 1 & 1 & 0 & 0 & 0 & 0 \\
  1 & 0 & 1 & 1 & 0 & 1 & 0 & 0 & 0 \\
  1 & 0 & 0 & 1 & 0 & 0 & 1 & 0 & 0 \\
  1 & 1 & 0 & 1 & 1 & 0 & 1 & 1 & 0 \\
  1 & 0 & 1 & 1 & 0 & 1 & 1 & 0 & 1 \\
  0 & 0 & 0 & 1 & 0 & 0 & 0 & 0 & 0 \\
  0 & 0 & 0 & 1 & 1 & 0 & 0 & 0 & 0 \\
  0 & 0 & 0 & 1 & 0 & 1 & 0 & 0 & 0 \\
  0 & 0 & 0 & 1 & 0 & 0 & 1 & 0 & 0 \\
  0 & 0 & 0 & 1 & 1 & 0 & 1 & 1 & 0 \\
  0 & 0 & 0 & 1 & 0 & 1 & 1 & 0 & 1 
  \end{array}
  \right), \hspace{0.2cm}
  \bflambda_{r}   = \left( \begin{array}{l}
  \lambda_{1}^{Y} \\ \lambda_{11}^{XY} \\ \lambda_{21}^{XY} \\ \lambda \\ \lambda_{1}^{X} \\  \lambda_{2}^{X} \\ \lambda_{1}^{Z} \\ \lambda_{11}^{XZ} \\ \lambda_{21}^{XZ} 
  \end{array}
  \right), \hspace{0.2cm}
  \bfT_{r} = \left( \begin{array}{llllllllll}
  1 & 0 & 0 & 0 & 0 & 0 & 0 & 0 & 0 \\
  0 & 1 & 0 & 0 & 0 & 0 & 0 & 0 & 0 \\
  0 & 0 & 1 & 0 & 0 & 0 & 0 & 0 & 0 
  \end{array}
  \right)
  \]
 \end{tiny}

The $g$-prior, 
\[
\bflambda \sim N(\bfm_{\lambda}, g \Sigma_{\lambda}) \equiv 
N((\mbox{log}(\bar{n}),0,\dots,0)^{\top}, \frac{g n_{ll}}{N} (X_{ll}^{\top} X_{ll})^{-1}),
\]
translates to, 
\[
\bflambda_{r} \sim N(\bfm_{\lambda_{r}}, g \Sigma_{\lambda_{r}}) \equiv 
N((0,\dots,0,\mbox{log}(\bar{n}),0,\dots,0)^{\top}, \frac{g n_{ll}}{N} (X_{rll}^{\top} X_{rll})^{-1}),
\]
where $\mbox{log}(\bar{n})$ is the $(n_{\beta}+1)$-th element in the mean vector. Then,
\[
E(\bfbeta)=E(\bfT_{r} \bflambda_{r}) = \bfT_{r} E(\bflambda_{r}) = \left( \begin{array}{ll}
    \bfI & \bf0  
  \end{array}
  \right) \times \bfmu_{\lambda_{r}}
 = \bf0. 
\]
Furthermore,
\[
\mbox{Var}(\bfbeta) = g \bfT_{r} \Sigma_{\lambda_{r}} \bfT_{r}^{\top} 
= \frac{g n_{ll}}{N} \bfT_{r} (X_{rll}^{\top} X_{rll})^{-1} \bfT_{r}^{\top} .
\]

From (A.1),
\begin{eqnarray}
 (X_{rll}^{\top} X_{rll})^{-1} &=& \left( \begin{array}{llll}
    X_{lt}^{* \top} X_{lt}^{*}  & X_{lt}^{* \top} X_{ll-lt}  \\
    X_{ll-lt}^{\top} X_{lt}^{*}  & X_{ll-lt}^{\top} X_{ll-lt}+X_{ll-lt}^{\top} X_{ll-lt}
  \end{array}
  \right)^{-1} \nonumber \\
&=& \left( \begin{array}{llll}
    X_{lt}^{* \top} X_{lt}^{*}  & X_{lt}^{* \top} X_{ll-lt}  \\
    X_{ll-lt}^{\top} X_{lt}^{*}  & 2 X_{ll-lt}^{\top} X_{ll-lt}
  \end{array}
  \right)^{-1}. \nonumber
\end{eqnarray}

From Lutkepohl (1996, p.147), the submatrix $H$ that is formed by the first $n_{\beta}$ rows and columns of $(X_{rll}^{\top} X_{rll})^{-1}$ is,
\[
H=(X_{lt}^{*\top} X_{lt}^{*})^{-1} 
\]
\[
+ (X_{lt}^{* \top} X_{lt}^{*})^{-1} X_{lt}^{* \top} X_{ll-lt} 
[X_{ll-lt}^{\top} (2 \bfI - X_{lt}^{*}(X_{lt}^{* \top} X_{lt}^{*})^{-1} X_{lt}^{* \top}) X_{ll-lt}]^{-1} 
\]
\[
\times X_{ll-lt}^{\top} X_{lt}^{*} (X_{lt}^{* \top} X_{lt}^{*})^{-1}. 
\]
Now, $P_{lt}\equiv X_{lt}^{*}(X_{lt}^{* \top} X_{lt}^{*})^{-1} X_{lt}^{* \top}$ is the projection matrix for $X_{lt}^{*}$. It is straightforward to verify that for a projection matrix $P_{lt}$ and a constant $c$,
\[
(c\bfI -P_{lt}) \times \left(\frac{1}{c}\bfI +\frac{1}{c(c-1)} P_{lt} \right) = \bfI.
\]

Therefore, $(2 \bfI - P_{lt})=(0.5 \bfI + 0.5 P_{lt})^{-1}$, and consequently, 
\[
H= (X_{lt}^{* \top} X_{lt}^{*})^{-1} + (X_{lt}^{* \top} X_{lt}^{*})^{-1} X_{lt}^{* \top} X_{ll-lt} 
[X_{ll-lt}^{\top} (0.5 \bfI + 0.5 P_{lt})^{-1} X_{ll-lt}]^{-1} 
\]
\[
\times X_{ll-lt}^{\top} X_{lt}^{*} (X_{lt}^{* \top} X_{lt}^{*})^{-1}.
\]

$X_{ll-lt}$ is a square matrix of full rank. If $X_{ll-lt}$ was not full rank, then some of its columns would be linearly dependent. In turn, some of the columns of 
$\left( \begin{array}{l}
    X_{ll-lt} \\
    X_{ll-lt}  
  \end{array}
  \right)$
would be linearly dependent, implying the same for columns of $X_{rll}$ [see equation (A.1)]. This is not possible as $X_{rll}$ is a design matrix of full rank. Thus, $X_{ll-lt}^{-1}$ exists and,
\begin{eqnarray}
H &=& (X_{lt}^{* \top} X_{lt}^{*})^{-1} \nonumber \\
&+& (X_{lt}^{* \top} X_{lt}^{*})^{-1} X_{lt}^{* \top} X_{ll-lt} 
[X_{ll-lt}^{-1} (0.5 \bfI + 0.5 P_{lt}) X_{ll-lt}^{\top^{-1}}] 
X_{ll-lt}^{\top} X_{lt}^{*} (X_{lt}^{* \top} X_{lt}^{*})^{-1} \nonumber \\
&=& (X_{lt}^{* \top} X_{lt}^{*})^{-1} + (X_{lt}^{* \top} X_{lt}^{*})^{-1} X_{lt}^{* \top}
(0.5 \bfI + 0.5 P_{lt})  
X_{lt}^{*} (X_{lt}^{* \top} X_{lt}^{*})^{-1} \nonumber \\
&=& (X_{lt}^{* \top} X_{lt}^{*})^{-1} + 0.5 (X_{lt}^{* \top} X_{lt}^{*})^{-1} + 0.5 (X_{lt}^{* \top} X_{lt}^{*})^{-1} \nonumber \\
&=& 2 (X_{lt}^{* \top} X_{lt}^{*})^{-1} \nonumber \\
&=& 2 (j_2 \times \hdots \times j_q  X_{lt}^{\top} X_{lt})^{-1} \nonumber 
\end{eqnarray}
Therefore, 
\begin{eqnarray}
\mbox{Var}(\bfbeta) &=& \frac{g n_{ll}}{N} \bfT_{r} (X_{rll}^{\top} X_{rll})^{-1} \bfT_{r}^{\top} \nonumber \\
&=& \frac{g n_{ll}}{N} \left( \begin{array}{ll} \bfI & \bf0  \end{array} \right) 
(X_{rll}^{\top} X_{rll})^{-1} 
\left( \begin{array}{l} \bfI \\ \bf0  \end{array} \right)  
\nonumber \\
&=& \frac{2  g  2 j_2 \times \hdots \times j_q  n_{lt}}{N j_2 \times \hdots \times j_q} (X_{lt}^{\top} X_{lt})^{-1} \nonumber \\
&=& \frac{4 g n_{lt}}{N} (X_{lt}^{\top} X_{lt})^{-1} \nonumber 
\end{eqnarray}
Thus, 
\[
\bfbeta \sim N({\bf 0} , \frac{4 g n_{lt}}{N} (X_{lt}^{\top} X_{lt})^{-1}),
\]
which is the $g$-prior for the parameters of a logistic regression, as described in Section 2. This completes the proof. 

\underline{Placing a flat prior on the Intercept:}  Assume that a flat prior is placed on the intercept of the log-linear model, after the design matrix has been centered to induce orthogonality between the intercept and the factors that form the contingency table. This does not alter the prior on the parameters of the corresponding logistic regression model. The proof follows along the lines of the proof of Theorem 1, if we express the parameters of the logistic regression model as $\bfbeta=\bfT_{r-1} \bflambda_{r-1}$, where $\bfT_{r-1}$ denotes matrix $\bfT_{r}$ without the first column with all elements zero, and $\bflambda_{r-1}$ denotes the vector of parameters $\bflambda_{r}$ without the intercept $\lambda$. The proof proceeds as above, replacing $X_{rll}$ with $X_{rll-1}$, where $X_{rll-1}$ is the former matrix without the column with all elements one. It is also required to replace $X_{ll-lt}$ with $X_{ll-lt-1}$, where $X_{ll-lt-1}$ is the former matrix without the column with all elements one.

\vspace{0.2cm}
{\it {\bf Proof of Theorem 2:}} The proof utilizes quantities defined earlier in Section 3 and in the proof of Theorem 1. First, we will show that, asymptotically, the posterior variance of $\bfbeta$ is identical to the posterior variance of the elements of $\bflambda$ that correspond to $\bfbeta$. Then, we will do the same for the posterior means. 

Consider a vector of cell counts $\bfn=\{n_1,\hdots,n_{ll} \}$, and the log-linear model $\mbox{log}(\bfmu) = X_{ll} \bflambda$. Then, asymptotically, 
\begin{eqnarray}
\mbox{Var}(\bflambda | \bfn) &\simeq& [g^{-1} \Sigma_{\lambda}^{-1} + {\cal I}(\hat{\bflambda})]^{-1} \nonumber \\
&=& \left[\frac{N}{g n_{ll}} X_{ll}^{\top} X_{ll} + X_{ll}^{\top} {\cal V}(\hat{\bflambda}) X_{ll} \right]^{-1}, \nonumber
\end{eqnarray}
where $\hat{\bflambda}$ denotes the maximum likelihood estimate (MLE). After rearranging the rows and columns of $X_{ll}$, consider the log-linear model with linear predictor $X_{rll} \bflambda_{r}$, for cell counts $\bfn_{r}$, where $\bfn_{r}$ is $\bfn$ rearranged to correspond to $X_{rll}$.  Now,

\begin{eqnarray}
\mbox{Var}(\bflambda_{r} | \bfn_{r}) &\simeq& [g^{-1} \Sigma_{\lambda_{r}}^{-1} + {\cal I}(\hat{\bflambda}_{r})]^{-1} \nonumber \\
&=& \left[\frac{N}{g n_{ll}} X_{rll}^{\top} X_{rll} + X_{rll}^{\top} {\cal V}(\hat{\bflambda}_{r}) X_{rll} \right]^{-1} \nonumber \\
&=& \left[X_{rll}^{\top} \left( \frac{N}{g n_{ll}} + {\cal V}(\hat{\bflambda_{r}}) \right) X_{rll} \right]^{-1} \nonumber \\
&=& \left[ 
\left(
\left( \begin{array}{llll}
   \frac{N}{g n_{ll}} \bfI+ {\cal V}_{1} {\cal V}_{2}  &  \bf0 \\
    \bf0  &  \frac{N}{g n_{ll}} + {\cal V}_{2}
  \end{array}
  \right)^{1/2} 
\left( \begin{array}{llll}
    X_{lt}^{*}  & X_{ll-lt}  \\
    \bf0   & X_{ll-lt}
  \end{array}
  \right) 
	\right)^{\top} \right.  \nonumber \\
	&\times&
\left. \left( \begin{array}{llll}
   \frac{N}{g n_{ll}} \bfI + {\cal V}_{1} {\cal V}_{2}  &  \bf0 \\
    \bf0  &  \frac{N}{g n_{ll}} + {\cal V}_{2}
  \end{array}
  \right)^{1/2} 
\left( \begin{array}{llll}
    X_{lt}^{*}  & X_{ll-lt}  \\
    \bf0   & X_{ll-lt}
  \end{array}
  \right) \right]^{-1}. \nonumber
\end{eqnarray}

${\cal V}_{1}$ denotes a diagonal matrix with non-zero elements $\mbox{exp}(X_{lt(i)}^{*} (\bfT_{r} \hat{\bflambda}_{r}))$, $i=1,\hdots,n_{ll}/2$. ${\cal V}_{2}$ denotes a diagonal matrix with non-zero elements $\mbox{exp}(X_{ll-lt(i)} \hat{\bflambda}_{ll-lt})$, $i=1,\hdots,n_{ll}/2$, where $\hat{\bflambda}_{ll-lt}$ denotes the MLE for $\bflambda_{r} \setminus \bfT_{r} \bflambda_{r}$. Now,
\[
\mbox{Var}(\bflambda_{r} | \bfn_{r}) \simeq 
\left( \begin{array}{llll}
    X_{lt}^{* \top} A_{12} X_{lt}^{*}  & X_{lt}^{* \top} A_{12} X_{ll-lt}  \\
    X_{ll-lt}^{\top} A_{12} X_{lt}^{*}  & X_{ll-lt}^{\top} (A_{12}+A_{2}) X_{ll-lt}
  \end{array}
  \right)^{-1} ,
\]
where, $A_{12}=\frac{N}{g n_{ll}} \bfI+ {\cal V}_{1} {\cal V}_{2} $ and $A_2=\frac{N}{g n_{ll}} \bfI+{\cal V}_{2}$. From Lutkepohl (1996, p.147), the submatrix $H$ that is formed by the first $n_{\beta}$ rows and columns of $\mbox{Var}(\bflambda_{r} | \bfn_{r})$ is,
\[
H=(X_{lt}^{*\top} A_{12} X_{lt}^{*})^{-1} + (X_{lt}^{* \top} A_{12} X_{lt}^{*})^{-1} X_{lt}^{* \top} A_{12} X_{ll-lt}
\]
\[
\times  
[X_{ll-lt}^{\top} (A_{12}+A_{2}) X_{ll-lt}  - 
  X_{ll-lt}^{\top} A_{12}  X_{lt}^{*}(X_{lt}^{* \top} A_{12} X_{lt}^{*})^{-1} X_{lt}^{* \top} A_{12} X_{ll-lt}]^{-1} 
\]
\[
\times X_{ll-lt}^{\top} A_{12} X_{lt}^{*} (X_{lt}^{* \top} A_{12} X_{lt}^{*})^{-1}
\]
\[
=(X_{lt}^{*\top} A_{12} X_{lt}^{*})^{-1} 
\]
\[
+   
(X_{lt}^{* \top} A_{12} X_{lt}^{*})^{-1} X_{lt}^{* \top} A_{12} [(A_{12}+A_{2})  - 
 A_{12}  X_{lt}^{*}(X_{lt}^{* \top} A_{12} X_{lt}^{*})^{-1} X_{lt}^{* \top} A_{12}]^{-1} 
\]
\[
\times A_{12} X_{lt}^{*} (X_{lt}^{* \top} A_{12} X_{lt}^{*})^{-1}
\]
\[
=(X_{lt}^{*\top} A_{12} X_{lt}^{*})^{-1} 
\]
\[
+   
(X_{lt}^{* \top} A_{12} X_{lt}^{*})^{-1} X_{lt}^{* \top} A_{12} 
[(\bfI +A_{12}^{-1} A_{2}) -  X_{lt}^{*}(X_{lt}^{* \top} A_{12} X_{lt}^{*})^{-1} X_{lt}^{* \top} A_{12}]^{-1} 
\]
\[
\times X_{lt}^{*} (X_{lt}^{* \top} A_{12} X_{lt}^{*})^{-1}
\]
\[
=(X_{lt}^{*\top} A_{12} X_{lt}^{*})^{-1} 
\]
\[
+   
(X_{lt}^{* \top} A_{12} X_{lt}^{*})^{-1} X_{lt}^{* \top} A_{12} 
[\bfI - (\bfI +A_{12}^{-1} A_{2})^{-1} X_{lt}^{*}(X_{lt}^{* \top} A_{12} X_{lt}^{*})^{-1} X_{lt}^{* \top} A_{12}]^{-1} 
\]
\[
\times X_{lt}^{*} (X_{lt}^{* \top} A_{12} X_{lt}^{*})^{-1}
\]
From Lutkepohl (1996, p.29, line 6), the expression above simplifies to,
\begin{eqnarray}
H &=& (X_{lt}^{*\top} A_{12} X_{lt}^{*} - X_{lt}^{*\top} A_{12} (\bfI +A_{12}^{-1} A_{2})^{-1} X_{lt}^{*})^{-1} \nonumber \\
&=& [X_{lt}^{*\top} (A_{12} - A_{12} (\bfI +A_{12}^{-1} A_{2})^{-1}) X_{lt}^{*}]^{-1}. \nonumber 
\end{eqnarray}
Within the Bayesian framework a large sample $(N \rightarrow \infty)$ will swamp the prior distribution, rendering it irrelevant for deriving posterior inferences (O'Hagan and Forster 2004). This can be viewed as equivalent to considering a flat non-informative prior, in our case assuming that $g \rightarrow \infty$. For a sample size large enough to justify ignoring the contribution of the prior distribution in $\mbox{Var}(\bflambda | \bfn)$, i.e. assuming that $A_{12}={\cal V}_{1} {\cal V}_{2} $ and $A_2={\cal V}_{2}$, asymptotically,  
\begin{eqnarray}
H &\simeq& [X_{lt}^{*\top} ({\cal V}_{1} {\cal V}_{2} - {\cal V}_{1} {\cal V}_{2} 
(\bfI +{\cal V}_{1}^{-1} {\cal V}_{2}^{-1} {\cal V}_{2})^{-1}) X_{lt}^{*}]^{-1} \nonumber \\
&=& [X_{lt}^{*\top} ({\cal V}_{1} {\cal V}_{2} - {\cal V}_{1}^2 {\cal V}_{2} 
(\bfI +{\cal V}_{1})^{-1}) X_{lt}^{*}]^{-1} \nonumber \\
&=& [X_{lt}^{*\top} [({\cal V}_{1} {\cal V}_{2} (\bfI + {\cal V}_{1}) - {\cal V}_{1}^2 {\cal V}_{2})  
(\bfI +{\cal V}_{1})^{-1}] X_{lt}^{*}]^{-1} \nonumber \\
&=& [X_{lt}^{*\top} ({\cal V}_{1} {\cal V}_{2} (\bfI +{\cal V}_{1})^{-1}) X_{lt}^{*}]^{-1} \nonumber \\
&=& [X_{lt}^{\top} ({\cal V}_{1,reduced} (\bfI +{\cal V}_{1,reduced})^{-1} 
 [{\cal V}_{2,1}+{\cal V}_{2,2} + \hdots +{\cal V}_{2,(j_1-1)\times j_2\times \hdots \times j_q}] X_{lt}]^{-1} \nonumber
\end{eqnarray}

${\cal V}_{1,reduced}$ denotes a diagonal matrix with elements $\mbox{exp}(X_{lt(i)} (\bfT_{r} \hat{\bflambda}_{r}))$, $i=1,\hdots,n_{lt}$. ${\cal V}_{2,k}$, $k=1,\hdots, (j_1-1)\times j_2\times \hdots \times j_q$, denotes a diagonal matrix with elements $\mbox{exp}(X_{ll-lt(n_{lt}(k-1)+i)} \hat{\bflambda}_{ll-lt})$. This expression simplifies as $q$ becomes smaller, i.e. the fewer times $X_{lt}$ is contained within $X_{lt}^{*}$. For example, when $X_{lt}^{*}=X_{lt}$, i.e. when $q=1$ and all factors other than $Y$ remain in the logistic regression, ${\cal V}_{1,reduced}={\cal V}_{1}$. 

We now utilize the standard result (see, for example, Rohatgi 1976, p.200) that, asymptotically, the Binomial distribution 
$Bin(t_i, \frac{\mbox{exp}(X_{lt(i)}^{*} (\bfT_{r} \bflambda_{r}))}{1+\mbox{exp}(X_{lt(i)}^{*} (\bfT_{r} \bflambda_{r}))})$ of a data point $t_{i} y_{i}$, $i=1,\hdots,n_{lt}$, can be approximated by a Poisson distribution 
$Poisson(t_i \frac{\mbox{exp}(X_{lt(i)}^{*} (\bfT_{r} \bflambda_{r}))}{1+\mbox{exp}(X_{lt(i)}^{*} (\bfT_{r} \bflambda_{r}))})$. The Binomial observation $t_i-t_i \times y_i$ is formed by adding $(j_1-1)\times j_2\times \hdots \times j_q$ independent Poisson cell counts. Considering the Poisson log-linear model, $t_i-t_i y_i$ follows the Poisson distribution,
\[
Poisson(\mbox{exp}(X_{ll-lt(i)} \hat{\bflambda}_{ll-lt})+\hdots + 
\mbox{exp}(X_{ll-lt(n_{lt}((j_1-1)\times j_2\times \hdots \times j_q-1)+i)} \hat{\bflambda}_{ll-lt})).
\]
Therefore, approximately, 
\[
t_i \frac{1}{1+\mbox{exp}(X_{lt(i)} (\bfT_{r} \hat{\bflambda}_{r}))}
\]
\[
\simeq \mbox{exp}(X_{ll-lt(i)} \hat{\bflambda}_{ll-lt})+\hdots + 
\mbox{exp}(X_{ll-lt(n_{lt}((j_1-1)\times j_2\times \hdots \times j_q-1)+i)} \hat{\bflambda}_{ll-lt}). \hspace{0.5cm} \mbox{(B.1)}
\]

In matrix notation, we can now write that, asymptotically, 
\begin{eqnarray}
\mbox{Var}(\bfT_{r} \bflambda_{r} | \bfn_{r}) &=& \bfT_{r} (\mbox{Var}(\bflambda_{r} | \bfn_{r}) ) \bfT_{r}^{\top} \nonumber \\
&=& \left( \begin{array}{ll} \bfI & \bf0  \end{array} \right) 
(\mbox{Var}(\bflambda_{r} | \bfn_{r}) ) 
\left( \begin{array}{l} \bfI \\ \bf0  \end{array} \right)  
\nonumber \\
&\simeq& [X_{lt}^{\top} ( \bft {\cal V}_{1,reduced} (\bfI + {\cal V}_{1,reduced})^{-2}) X_{lt}]^{-1} \nonumber \\
&=& (X_{lt}^{\top} {\cal V}_{logistic} X_{lt})^{-1}     \nonumber 
\end{eqnarray}
where, $\bft$ is a diagonal matrix with diagonal elements the number of trials $t_i$, and ${\cal V}_{logistic}$ has diagonal elements $t_i \mbox{exp}\{ X_{lt(i)} \hat{\bfbeta} \} \mbox{exp}\{1 + X_{lt(i)} \hat{\bfbeta} \}^{-2}$, $i=1,\dots,n_{lt}$. $(X_{lt}^{\top} {\cal V}_{logistic} X_{lt})^{-1}$ is, asymptotically, the posterior variance of $\bfbeta$ when the logistic regression is fitted directly, and thus we have shown that the posterior variance of $\bfbeta$ is identical to the posterior variance of the elements of $\bflambda$ that correspond to $\bfbeta$.

We will now show that, asymptotically, the posterior mean $E(\bfbeta | \bft, \bfy)$ is the posterior mean of the elements of $\bflambda$ that correspond to $\bfbeta$. For a sample large enough to justify ignoring the contribution of the prior in (1), we obtain that, $E(\bflambda | \bfn)\simeq {\cal I}(\hat{\lambda})^{-1} {\cal I}(\hat{\lambda}) \hat{\lambda} = \hat{\lambda}$. Similarly, $E(\bfbeta | \bft, \bfy)\simeq \hat{\bfbeta}$. 
Therefore, $E(\bfT_{r} \bflambda_{r} | \bfn)\simeq \bfT_{r} \hat{\bflambda}_{r}$, and it is sufficient to show that $\hat{\bfbeta}=\bfT_{r} \hat{\bflambda}_{r}$. Closed form expressions for the maximum likelihood estimators of the parameters of a generalized linear model do not exist. As a result, we will base the derivation of this result on the Iterative Re-weighed Least Squares (IRLS) algorithm. This is the standard procedure for maximizing the likelihood when a generalized model is fitted. See Wood (2006) for more  details. For a linear predictor $X_d \bfgamma$ this iterative process is based on the formula,
\[
\bfgamma^{it+1}=\bfgamma^{it} + (X_d^{\top} {\cal V}(\bfgamma^{it}) X_d)^{-1} X_d^{\top} {\cal V}(\bfgamma^{it}) \bfzeta^{it}.
\]
For a log-linear model, $\bfzeta^{it}$ is denoted by $\bfzeta^{it}_{log-linear}$, and its $i$-th element, $i=1,\hdots, n_{ll}$, is, 
\[
\zeta_{log-linear(i)}=\frac{n_i}{\mbox{exp}(X_{rll(i)} \bflambda_{r}^{it})} - 1.
\]
For a logistic regression model, $\bfzeta^{it}$ is denoted by $\bfzeta^{it}_{logistic}$, and its $i$-th element, $i=1,\hdots, n_{lt}$, is, 
\[
\zeta_{logistic(i)}=\frac{t_i y_i (1+\mbox{exp}(X_{lt} \bfbeta^{it})) - t_i \mbox{exp}(X_{lt} \bfbeta^{it}) }
{t_i} \frac{1+\mbox{exp}(X_{lt} \bfbeta^{it})}{\mbox{exp}(X_{lt} \bfbeta^{it})}. 
\]
For the log-linear model, the IRLS procedure is written as, 
\[
\bflambda_{r}^{it+1}=\bflambda_{r}^{it} + (X_{rll}^{\top} {\cal V}_{log-linear}(\bflambda_{r}^{it}) X_{rll})^{-1} X_{rll}^{\top} {\cal V}_{log-linear}(\bflambda_{r}^{it}) \bfzeta^{it}_{log-linear},
\]
where ${\cal V}_{log-linear}$ is a diagonal matrix with diagonal elements $\mbox{exp}\{ X_{rll(i)} \hat{\bflambda}_{r} \}$, $i=1,\dots,n_{ll}$. Algebraic operations similar to the ones carried out earlier show that $(X_{rll}^{\top} {\cal V}_{log-linear}(\bflambda^{it}) X_{rll})^{-1}$ partitions as,
\[
\left( \begin{array}{llll}
    (X_{lt}^{\top} {\cal V}_{logistic} X_{lt})^{-1}  & 
			- [X_{lt}^{* \top} {\cal V}_{1} {\cal V}_{2} X_{lt}^{*}]^{-1} X_{lt}^{* \top} {\cal V}_{1}  
\times [{\cal V}_{1} + \bfI - {\cal V}_{1} {\cal V}_{2} X_{lt}^{*} \\
&  \times[X_{lt}^{* \top} {\cal V}_{1} {\cal V}_{2} X_{lt}^{*}]^{-1} X_{lt}^{* \top} {\cal V}_{1}]^{-1} X_{ll-lt}^{\top -1} \\
   \Omega_1   & \Omega_2
  \end{array}
  \right) ,
\]

where $\Omega_1$ and $\Omega_2$ are matrices not relevant to this proof. Furthermore, \\ $X_{rll}^{\top} {\cal V}_{log-linear}(\bflambda^{it}_{r})$ partitions as, 
\[
\left( \begin{array}{llll}
    X_{lt}^{* \top} {\cal V}_{1} {\cal V}_{2}  & \bf0 \\
    X_{ll-lt}^{\top} {\cal V}_{1} {\cal V}_{2}  & X_{ll-lt}^{\top} {\cal V}_{2}
  \end{array}
  \right). 
\]

For the log-linear model, we write $\bfzeta_{log-linear}=(\bfzeta_{lt}^{*\top} \bfzeta_{ll-lt}^{\top})^{\top}$, where $\bfzeta_{lt}^{*}$ corresponds to the first $n_{ll}/2$ rows of $X_{rll}$. Now, the first  $n_{\beta}$ elements of \\
$(X_{rll}^{\top} {\cal V}_{log-linear}(\bflambda^{it}) X_{rll})^{-1} 
X_{rll}^{\top} {\cal V}_{log-linear}(\bflambda^{it}_{r}) \bfzeta_{log-linear}$, i.e. the ones that correspond to the logistic regression parameters, are given by, 
\[
(X_{lt}^{\top} {\cal V}_{logistic} X_{lt})^{-1} X_{lt}^{* \top} {\cal V}_{1} {\cal V}_{2} \bfzeta_{lt}^{*}
\]
\[
- [X_{lt}^{* \top}  {\cal V}_{1} {\cal V}_{2} X_{lt}^{*}]^{-1} X_{lt}^{* \top} {\cal V}_{1} \times
[{\cal V}_{1} + \bfI - {\cal V}_{1} {\cal V}_{2} X_{lt}^{*} 
(X_{lt}^{* \top}  {\cal V}_{1} {\cal V}_{2} X_{lt}^{*})^{-1} X_{lt}^{*\top} {\cal V}_{1}]^{-1} \times
\]
\[
[{\cal V}_{1} {\cal V}_{2} \bfzeta_{lt}^{*} + {\cal V}_{2} \bfzeta_{ll-lt}].
\]
The $i$-th element of $\bfzeta_{lt}^{*}$, $i=1,\hdots,n_{ll}/2$, is,   
\[
\zeta_{lt(i)}=\frac{n_i}{\mbox{exp}(X_{lt(i)} \bfT_{r} \bflambda_{r}^{it}) \mbox{exp}(X_{ll-lt(i)} \bflambda_{ll-lt}^{it})} -1. 
\] 
The $i$-th element of $\bfzeta_{ll-lt}$, $i=1,\hdots,n_{ll}/2$, is,  
\[
\zeta_{ll-lt(i)}=\frac{t_i - n_i}{ \mbox{exp}(X_{ll-lt(i)} \bflambda_{ll-lt}^{it})} -1. 
\]
It is straightforward to show that $[{\cal V}_{1} {\cal V}_{2} \bfzeta_{lt}^{*} + {\cal V}_{2} \bfzeta_{ll-lt}]$ is, approximately, a vector of zeros. To show this, consider, without loss of generality, the $i$-th element of this vector, 
\[
\mbox{exp}(X_{lt(i)} \bfT_{r} \bflambda_{r}^{it}) \mbox{exp}(X_{ll-lt(i)} \bflambda_{ll-lt}^{it}) \times 
[\frac{n_i}{\mbox{exp}(X_{lt(i)} \bfT_{r} \bflambda_{r}^{it}) \mbox{exp}(X_{ll-lt(i)} \bflambda_{ll-lt}^{it})} -1] 
\]
\[
+ \mbox{exp}(X_{ll-lt(i)} \bflambda_{ll-lt}^{it}) \times [\frac{t_i - n_i}{ \mbox{exp}(X_{ll-lt(i)} \bflambda_{ll-lt}^{it})} -1]
\]
\[
= t_i - \mbox{exp}(X_{ll-lt(i)} \bflambda_{ll-lt}^{it}) \times [1+\mbox{exp}(X_{lt(i)} \bfT_{r} \bflambda_{r}^{it})].
\]
Due to the Poisson approximation to the Binomial distribution, 
\[
\mbox{exp}(X_{ll-lt(i)} \bflambda_{ll-lt}^{it}) \simeq t_i \frac{1}{1+\mbox{exp}(X_{lt(i)} \bfT_{r} \bflambda_{r}^{it})}.
\]
Thus, the elements of vector $[{\cal V}_{1} {\cal V}_{2} \bfzeta_{lt} + {\cal V}_{2} \bfzeta_{ll-lt}]$ are all zero, and
the first  $n_{\beta}$ elements of 
$(X_{rll}^{\top} {\cal V}_{log-linear}(\bflambda^{it}) X_{rll})^{-1} 
X_{rll}^{\top} {\cal V}_{log-linear}(\bflambda^{it}) \bfzeta_{log-linear}$ are approximately equal to,
\[
(X_{lt}^{\top} {\cal V}_{logistic} X_{lt})^{-1} X_{lt}^{* \top} {\cal V}_{1} {\cal V}_{2} \bfzeta_{lt}^{*}
\]
\[
= (X_{lt}^{\top} {\cal V}_{logistic} X_{lt})^{-1} X_{lt}^{\top} {\cal V}_{1,reduced} 
({\cal V}_{2,1} \hdots {\cal V}_{2,(j_1-1)\times j_2 \times \hdots \times j_q} ) \bfzeta_{lt}^{*}.
\]
Using the Poisson approximation to the Binomial distribution, for the $i$-th element of $\bfzeta_{lt}^{*}$, and  
assuming without any loss of generality that $i<n_{lt}$, 
\[
\zeta_{lt(i)}^{*}\simeq \frac{n_i}{\mbox{exp}(X_{rll(i)} \bflambda_{r}^{it})} - 1
= \frac{n_i}{\mbox{exp}(X_{lt(i)} \bfT_{r} \bflambda_{r}^{it}) t_i 
\frac{1}{1+\mbox{exp}(X_{lt(i)} \bfT_{r} \bflambda_{r}^{it})}} - 1
\]
\[
= \frac{n_i (1+\mbox{exp}(X_{lt(i)} \bfT_{r} \bflambda_{r}^{it})) - t_i \mbox{exp}(X_{lt(i)} \bfT_{r} \bflambda_{r}^{it})}
{t_i \mbox{exp}(X_{lt(i)} \bfT_{r} \bflambda_{r}^{it})}.
\]
Thus, 
\[
\zeta_{lt(i)}^{*}\simeq(1+\mbox{exp}(X_{lt(i)} \bfT_{r} \bflambda_{r}^{it}))^{-1} \zeta_{logistic(i)}. 
\]
Therefore, the updating step for $\bfT_{r} \bflambda_{r}$ is,
\[
\bfT_{r} \bflambda_{r}^{it+1}=\bfT_{r} \bflambda_{r}^{it} + 
(X_{lt}^{\top} {\cal V}_{logistic} X_{lt})^{-1} X_{lt}^{\top} 
\]
\[
\times {\cal V}_{1,reduced} (\bfI + {\cal V}_{1,reduced})^{-1} 
({\cal V}_{2,1} \hdots {\cal V}_{2,(j_1-1)\times j_2 \times \hdots \times j_q} )  
( \bfzeta_{logistic}^{it \top} \hdots \bfzeta_{logistic}^{it \top})^{\top} .
\]
\[
= \bfT_{r} \bflambda_{r}^{it} + 
(X_{lt}^{\top} {\cal V}_{logistic} X_{lt})^{-1} X_{lt}^{\top} 
\]
\[
\times {\cal V}_{1,reduced} (\bfI + {\cal V}_{1,reduced})^{-1} 
({\cal V}_{2,1} + \hdots + {\cal V}_{2,(j_1-1)\times j_2 \times \hdots \times j_q} )  
\bfzeta_{logistic}^{it} .
\]

If the logistic regression was to be fitted directly, obtaining the MLE would be based on the IRLS algorithm, 
\[
\bfbeta^{it+1}=\bfbeta^{it} + (X_{lt}^{\top} {\cal V}_{logistic}(\bfbeta^{it}) X_{lt})^{-1} X_{lt}^{\top} 
\times {\cal V}_{logistic}(\bfbeta^{it}) \bfzeta_{logistic}^{it}.
\]

By replacing the sum of the elements of the ${\cal V}_{2,k}$ matrices with the approximate values given in (B.1), we observe that, asymptotically, the updating step is the same for both $\bfT_{r} \bflambda_{r}$ and $\bfbeta$. Thus, if the starting point for $\bfT_{r} \bflambda_{r}$ is the same as the starting point for $\bfbeta$, the iterative algorithm would give the same MLE for the logistic regression parameters and the corresponding log-linear model parameters. The IRLS algorithm is robust to different starting values when the likelihood is not flat. Therefore, asymptotically, $\hat{\bfbeta}=\bfT_{r} \hat{\bflambda}_{r}$ and the proof is complete.

% BibTeX users please use one of
%\bibliographystyle{spbasic}      % basic style, author-year citations
%\bibliographystyle{spmpsci}      % mathematics and physical sciences
%\bibliographystyle{spphys}       % APS-like style for physics
%\bibliography{}   % name your BibTeX data base

% Non-BibTeX users please use
%\begin{thebibliography}{}
%
% and use \bibitem to create references. Consult the Instructions
% for authors for reference list style.
%
%\bibitem{RefJ}
% Format for Journal Reference
%Author, Article title, Journal, Volume, page numbers (year)
% Format for books
%\bibitem{RefB}
%Author, Book title, page numbers. Publisher, place (year)
% etc
%\end{thebibliography}

\parindent 0mm \parskip 0cm
\vspace{0.3cm}
\begin{large}
\noindent {\bf References}
\end{large}
\vspace{0.0cm}

\begin{list}{}
{\setlength{\itemsep}{0cm}
\setlength{\parsep}{0cm}
\setlength{\leftmargin}{0.5cm}
\setlength{\labelwidth}{0.5cm}
\setlength{\itemindent}{-0.5cm}
}

\item Agresti A (2002) Categorical data analysis. second ed. John Wiley and Sons, New Jersey

\item Bapat RB (2011) Graphs and Matrices. Springer. Hindustan Book Agency, New Delhi

\item Consonni G, Veronese P. (2008) Compatibility of prior specifications across linear models. Stat Sci 23:232-353

\item Dellaportas P, Forster JJ (1999) Markov chain Monte Carlo model determination for hierarchical and graphical log-linear models. Biometrika 86:615-633

\item Dellaportas P, Forster JJ, Ntzoufras I (2012) Joint specification of model space and parameter space prior distributions. Stat Sci 27:232-246

\item Edwards D, Havr\'anek T (1985) A fast procedure for model search in multi-dimensional contingency tables. Biometrika 72:339-351

\item Fouskakis D, Ntzoufras I, Draper D (2015) Power-expected-posterior priors for variable selection in Gaussian linear models. Bayesian Anal 10:75-107

\item Held L, Saban\`{e}s Bov\`{e} D, Gravestock I (2015) Approximate Bayesian model selection with the deviance statistic. Stat Sci 
\\ \verb http://www.imstat.org/sts/future_papers.html \verb Accessed 17 March 2016

\item Kass RE, Wasserman L (1995) A reference Bayesian test for nested hypotheses and its relationship to the Schwarz criterion. J Am Stat Assoc 90:928-934

\item Liang F, Paulo R, Molina G, Clyde MA, Berger JO (2008) Mixtures of g-priors for Bayesian variable selection. J Am Stat Assoc 103:410-423 

\item Lutkepohl H (1996) Handbook of matrices. John Wiley and Sons, Chichester

\item Mukhopadhyay M, Samantha T (2016) A mixture of g-priors for variable selection when
the number of regressors grows with the sample size. Test DOI 10.1007/s11749-016-0516-0

\item Ntzoufras I, Dellaportas P, Forster JJ (2003) Bayesian variable and link determination for generalized linear models. J Stat Plan Infer 111:165-180

\item Ntzoufras I (2009) Bayesian modelling using WinBugs. John Wiley and Sons, New Jersey

\item O'Hagan A (1995) Fractional Bayes factors for model comparison. J R Stat Soc Ser B 57:99-138 

\item O'Hagan A, Forster JJ (2004) Bayesian Inference. second ed. vol 2B of `Kendall's Advanced Theory of Statistics'. Arnold, London 

\item Overstall A, King R (2014a) A default prior distribution for contingency tables with dependent factor levels. Stat Methodol 16:90-99 

\item Overstall A, King R (2014b) conting: an R package for Bayesian analysis of complete and incomplete contingency tables. J Stat Softw 58:1-27 

\item Papathomas M, Richardson S (2016) Exploring dependence between categorical variables: benefits and limitations of using variable selection within Bayesian clustering in relation to log-linear modelling with interaction terms. J Stat Plan Infer 173:47-63 

\item Papathomas M, Dellaportas P, Vasdekis VGS (2011) A novel reversible jump algorithm for generalized linear models. Biometrika 98:231-236

\item Rohatgi VK (1976) An introduction to probability theory and mathematical statistics. John Wiley and Sons, New York

\item Saban\`{e}s Bov\`{e} D, Held L (2011) Hyper-g priors for generalized linear models. Bayesian Anal 6:387-410

\item Wang X, George GI (2007) Adaptive Bayesian criteria in variable selection for generalized linear models. Stat Sinica 17:667-690

\item Wood SN (2006) Generalized additive models. An introduction with R. Chapman and Hall/CRC, New York

\item Zellner A (1986) On assessing prior distributions and Bayesian regression analysis with g-prior distributions. In: Goel PK, Zellner A (eds) Bayesian Inference and Decision Techniques: Essays in Honor of Bruno de Finetti. North-Holland/Elsevier, pp 233-243

\end{list}

\newpage

\begin{small}
\begin{center}
\begin{table*}[!t]
\begin{small}
{\bf Table 1:} {Simulated data illustration. Credible intervals (CIs) for the relevant parameters of log-linear model (M2), plus the parameters of the corresponding  logistic regression (M3). 
}
\label{tab:1}
\par
\begin{tabular}{cccccccc}
\hline
\multicolumn{8}{c}{Log-linear model (M2), \hspace{0.8cm} $\mbox{log}(\bfmu) = YAB + YCD + YE + ABCDE$} \\
Y & YA  &  YB & YC & YD & YE & YAB & YCD   \\
(0.21,1.07) & (-0.57,0.26) & (-0.44,0.43) &  (-0.24,0.63) & (-0.38,0.50) & (-0.84,-0.27) & (-1.66,-0.50) &  (-2.01,-0.85)    \\
\hline
\multicolumn{8}{c}{Outcome is Y (M3), \hspace{0.8cm} $\mbox{logit}(\bfp) = AB+CD+E$} \\
Intercept & A & B & C & D & E & AB & CD    \\
(0.21,1.08) & (-0.58,0.27) & (-0.45,0.43) & (-0.23,0.61) & (-0.38,0.49) & (-0.84,-0.27) & (-1.66,-0.50) & (-2.00,-0.84)    \\

\hline 
\end{tabular}
\end{small}
\end{table*}
\end{center}
\end{small}

\begin{small}
\begin{center}
\begin{table*}[!h]
\begin{small}
{\bf Table 2:} {Simulated data illustration. Maximum, minimum, and quantiles for $t_i$, $i=1,\hdots, n_{lt}$, for each of the logistic regressions shown in Table 1.}
\label{tab:1}
\par
\begin{tabular}{cccccc}
\hline
Outcome & Minimum & 25\% Quantile & Median & 75\% Quantile & Maximum \\
\hline 
Y & 11 &  17 &   21 &   41.5 &  124   \\
A &  12 &   19 &   23 &   30 &  144  \\
B & 10 &   18 &   22.5 &   31 &  165 \\
C & 12 &   18.5 &   23 &   26.5 &  151 \\
D &  11 &   19.5 &   23 &   27.5 &  147 \\
E & 10 &   17.5 &   22 &   27 &  191 \\
\hline 
\end{tabular}
\end{small}
\end{table*}
\end{center}
\end{small}

\begin{small}
\begin{center}
\begin{table*}[!h]
\begin{small}
{\bf Table 3:} {Real data illustration. Relevant credible intervals for the parameters of log-linear model (M4) and the corresponding logistic regression model when A is treated as the outcome. Intervals are shown under the $g$-priors in Section 2 (g=N), and after considering a locally flat prior on the intercepts.}
\label{tab:2}
\par
\begin{tabular}{cccccccc}
\hline
\multicolumn{8}{c}{Log-linear model (M4), \hspace{0.2cm} $\mbox{log}(\bfmu) = AC + AD + AE + BCDEF$ \hspace{0.2cm} ($g$-prior in Section 2)} \\
A & AC & AD & AE & & &  &     \\
 (-0.59,-0.24) & (0.36,0.74) & (-0.56,-0.18) & (0.30,0.68) & & & &   \\
\hline
\multicolumn{8}{c}{Outcome is A (M5), \hspace{0.3cm} $\mbox{logit}(\bfp) = C+D+E$ \hspace{0.2cm} ($g$-prior in Section 2)} \\ 
Intercept & C & D & E & & &  &     \\
(-0.59,-0.24) & (0.37,0.74) & (-0.56,-0.18) & (0.30,0.68) & & & &   \\
\hline
\multicolumn{8}{c}{Log-linear model (M4), \hspace{0.2cm} $\mbox{log}(\bfmu) = AC + AD + AE + BCDEF$ \hspace{0.2cm} (flat prior on intercept)} \\
A & AC & AD & AE & & &  &     \\
 (-0.59,-0.24) & (0.35,0.76) & (-0.55,-0.19) & (0.29,0.67) & & & &   \\
\hline
\multicolumn{8}{c}{Outcome is A (M5), \hspace{0.3cm} $\mbox{logit}(\bfp) = C+D+E$ \hspace{0.2cm} (flat prior on intercept)} \\ 
Intercept & C & D & E & & &  &     \\
(-0.17,0.02) & (0.35,0.75) & (-0.56,-0.19) & (0.30,0.68) & & & &   \\
\hline
\end{tabular}
\end{small}
\end{table*}
\end{center}
\end{small}

\end{document}